\documentclass[prd,preprint,floatfix,nofootinbib,superscriptaddress,,showkeys]{revtex4} %,twocolumn
\pdfoutput=1
\usepackage[caption=false]{subfig}
\usepackage{amsmath}
\usepackage{amsfonts}
\usepackage{amssymb}
\usepackage{float}
\usepackage{color}
\usepackage{graphicx}
\usepackage{graphics}
\usepackage{hyperref}
\usepackage{adjustbox,array}
\usepackage{enumitem} 
\hypersetup{}
\usepackage[utf8x]{inputenc} 
\usepackage{epstopdf}
\usepackage{multirow}
\usepackage{slashed}
\usepackage{booktabs}
\usepackage{cancel}
\usepackage{mathrsfs}
\usepackage{csquotes}

\definecolor{rosso}{cmyk}{0,1,1,0.4}
\definecolor{rossos}{cmyk}{0,1,1,0.55}
\definecolor{rossoc}{cmyk}{0,1,1,0.2}
\definecolor{blu}{cmyk}{1,1,0,0.3}
\definecolor{blus}{cmyk}{1,1,0,0.6}
\definecolor{bluc}{cmyk}{1,1,0,0.1}
\definecolor{verde}{cmyk}{0.92,0,0.59,0.25}
\definecolor{verdec}{cmyk}{0.92,0,0.59,0.15}
\definecolor{verdes}{cmyk}{0.92,0,0.59,0.4}

\AtBeginDocument{
\heavyrulewidth=.08em
\lightrulewidth=.05em
\cmidrulewidth=.03em
\belowrulesep=.65ex
\belowbottomsep=0pt
\aboverulesep=.4ex
\abovetopsep=0pt
\cmidrulesep=\doublerulesep
\cmidrulekern=.5em
\defaultaddspace=.5em
}

\hypersetup{
 colorlinks=true,
 linkcolor=verdec,
 urlcolor=verdec,
 citecolor=verdec%gray
}

\newcommand{ \eq}[1]{Eq.~(\ref{#1})}

\newcommand{\gsim}{\gtrsim}
\newcommand{\lsim}{\lesssim}

\newcommand{\lf}{\left(}
\newcommand{\ri}{\right)}

\newcommand{\nn}{\nonumber}

\newcommand{\sqt}{\sqrt{2}}
\newcommand{\hf}{h_{\rm f}}
\newcommand{\rr}{{\gamma\gamma}}

\renewcommand{\lg}{\mathscr{L}} % Amplitude
\newcommand{\mco}{\mathcal{O}}

\newcommand{\br}{\mathcal{B}}
\newcommand{\hc}{{\rm H.c.}}

\newcommand{\tot}{{\rm tot}}

\newcommand{\fb}{{\;{\rm fb}}}
\newcommand{\ab}{{\;{\rm ab}}}

\newcommand{\gev}{{\;{\rm GeV}}}
\newcommand{\tev}{{\;{\rm TeV}}}

\newcommand{\beq}{\begin{equation}}
\newcommand{\eeq}{\end{equation}}
\newcommand{\bea}{\begin{eqnarray}}
\newcommand{\eea}{\end{eqnarray}}
\newcommand{\barr}{\begin{array}}
\newcommand{\earr}{\end{array}}
\newcommand{\bc}{\begin{center}}
\newcommand{\ec}{\end{center}}
\newcommand{\bit}{\begin{itemize}}
\newcommand{\eit}{\end{itemize}}
\newcommand{\ben}{\begin{enumerate}}
\newcommand{\een}{\end{enumerate}}
\newcommand{\al}{\alpha}
\newcommand{\bt}{\beta}

\newcommand{\Dt}{\Delta}

\newcommand{\sg}{\sigma}
\newcommand{\es}{\epsilon}
\newcommand{\kp}{\kappa}

\newcommand{\lmh}{\hat{\lambda}}

\newcommand{\gm}{\gamma}
\newcommand{\Gm}{\Gamma}

\newcommand{\hsm}{{h_{\rm SM}}}
\newcommand{\ch}{H^\pm}

\newcommand{\wpm}{W^\pm}

\newcommand{\mhsm}{m_{125}}
\newcommand{\mch}{M_{H^\pm}}
\newcommand{\mhh}{M_{H}}
\newcommand{\ma}{M_{A}}
\newcommand{\mach}{M_{A/H^\pm}}
\newcommand{\mhf}{m_{h_{\rm f}}}
\newcommand{\dma}{\Delta M_{A}}
\newcommand{\lmc}{\Lambda_{\rm c}}
\newcommand{\ca}{c_\alpha}
\newcommand{\sa}{s_\alpha}

\newcommand{\tb}{t_\beta}

\newcommand{\cb}{c_\beta}
\renewcommand{\sb}{s_\beta}

\newcommand{\cba}{c_{\beta-\alpha}}
\newcommand{\sba}{s_{\beta-\alpha}}
\newcommand{\ee}      {{e^+ e^-}}

\newcommand{\ttau}      {{\tau^+\tau^-}} %{{\tau\tau}} %

\newcommand{\ttop}      {{t\bar{t}}}

\newcommand{\bb}      {{b \bar{b}}}

\newcommand{\qq}      {{q \bar{q}}}

\newcommand{\elll}      {{\ell^+\ell^- }}

\begin{document}

\title{\color{verdes} Fermiophobic light Higgs boson \\
in the type-I two-Higgs-doublet model}
\author{Jinheung Kim}
\email{jinheung.kim1216@gmail.com}
\address{Department of Physics, Konkuk University, Seoul 05029, Republic of Korea}
\author{Soojin Lee}
\email{soojinlee957@gmail.com}
\address{Department of Physics, Konkuk University, Seoul 05029, Republic of Korea}
\author{Prasenjit Sanyal}
\email{prasenjit.sanyal@apctp.org}
\address{Asia Pacific Center for Theoretical Physics, Pohang 37673, Republic of Korea}
\author{Jeonghyeon Song}
\email{jhsong@konkuk.ac.kr}
\address{Department of Physics, Konkuk University, Seoul 05029, Republic of Korea}

\begin{abstract}
The null results in the new physics searches at the LHC do not exclude an intermediate-mass new particle if it is fermiophobic. Type-I in the two-Higgs-doublet model accommodates a fermiophobic light Higgs boson $h_{\rm f}$ if $\al=\pi/2$. The heavier \textit{CP}-even Higgs boson explains the observed Higgs boson at a mass of 125~GeV. We first obtain the still-valid parameter space satisfying the theoretical requirements, flavor-changing neutral currents in $B$ physics, the cutoff scale above 1 TeV, Higgs precision data, and the direct collider search bounds at high energy colliders. We also study the high energy scale behavior via the analysis of the renormalization group equations. An important result is that the fermiophobic type-I can maintain the stability of the scalar potential all the way up to the Planck scale if $m_{h_{\rm f}}$ is larger than half the observed Higgs boson mass. Since the parameter space is severely curtailed, especially for the high cutoff scale, the signal rates in the multi-photon states of new Higgs bosons at the LHC are well predicted. We suggest the  processes of $4\gamma+VV'$ ($V^{(\prime)}=Z, W^\pm$) as the golden discovery channels for the model since they enjoy an almost background-free environment and substantial selection efficiencies for four photons.
 \end{abstract}

\vspace{1cm}
\preprint{APCTP Pre2022--016}

\maketitle
\tableofcontents
%\flushbottom 

\section{Introduction}

The standard model (SM) of particle physics has achieved unprecedented success
in explaining almost all the experimental data at high energy colliders.
Nevertheless, we firmly believe that there is new physics beyond the SM (BSM)
because the SM does not have the answers for the naturalness problem, the fermion mass hierarchy, 
the origin of \textit{CP} violation in the quark sector, the baryogenesis, the non-zero neutrino masses, 
and the identity of dark matter.
In the hope that a new physics signal is just around the corner,
the ATLAS and CMS collaborations have tried hard to find a new particle,
but not succeeded so far.

The absence of a new signal in the current data set at the LHC 
is usually attributed to very heavy or very light masses of new particles. 
But there exists an intriguing alternative, a fermiophobic new particle.
If a new particle does not couple to the SM fermions,
its production via quark-antiquark annihilation is highly suppressed.
If the new particle is a color singlet as suggested in many BSM models,
the gluon fusion production through quark loops is also forbidden.
The infeasibility of the direct production of a fermiophobic particle naturally explains the absence of new signals 
at the LHC.
In this regard, there have been extensive theoretical studies on the fermiophobic BSM particle 
such as a Higgs boson~\cite{Akeroyd:1995hg,Akeroyd:1998ui,Akeroyd:1998dt,Barroso:1999bf,Brucher:1999tx,Akeroyd:2003bt,Akeroyd:2003xi,Akeroyd:2007yh,Arhrib:2008pw,Gabrielli:2012yz,Berger:2012sy,Gabrielli:2012hd,Cardenas:2012bg,Ilisie:2014hea,Delgado:2016arn,Mondal:2021bxa},
gauge boson~\cite{Donini:1997yu,Ohl:2008ri,Bramante:2011qc,Bach:2011jy,Chen:2014noa,Coleppa:2018fau,Navarro:2021sfb,Fu:2021uoo}, and 
unparticle~\cite{Thalapillil:2009ch}.
In the experimental side, a fermiophobic Higgs boson has been searched for in the diphoton mode at the LEP-2~\cite{DELPHI:2001pxd,DELPHI:2003hpv},
Tevatron~\cite{D0:2008srr,CDF:2009pne,D0:2011kow,CDF:2011wic},
and LHC~\cite{CMS:2012mua,ATLAS:2012yxc,CMS:2012xok,CMS:2013zma}.

Nevertheless criticism can follow that a BSM model with a fermiophobic particle
is introduced as an ad hoc just to evade the disappointing situation of no new signals at the electroweak scale. 
One way to value the BSM model is to investigate the high energy scale behavior through 
the analysis of the renormalization group equations (RGE).
If the model remains theoretically valid up to very high energy scale like the Planck scale,
it will become a better candidate for the UV theory.
The RGE analysis requires one specific BSM model.
Type I in the two-Higgs-doublet model (2HDM) is one of the most attractive BSM models
to offer a fermiophobic Higgs boson:
all the Yukawa couplings of a new \textit{CP}-even neutral Higgs boson $\varphi^0$ are the same in type I
so that a single condition can guarantee the fermiophobic nature of $\varphi^0$.
The observation of the SM-like Higgs boson at the LHC~\cite{Aad:2019mbh,CMS:2020xwi,ATLAS:2021vrm}
can be explained in two ways, 
the normal scenario where the lighter \textit{CP}-even Higgs boson is observed
and the inverted scenario where the heavier \textit{CP}-even Higgs boson is observed~\cite{Bernon:2015wef,Chang:2015goa,Jueid:2021avn,Lee:2022gyf}.
Considering the experimental searches for a light fermiophobic Higgs boson via diphoton mode,
we focus on the inverted scenario of type I in this letter.

The study of the high energy scale behavior of the fermiophobic type-I requires
the basic work of extracting all the viable parameter points that satisfy the theoretical and experimental constraints,
via random scanning of the entire parameter space.
In the literature, the Higgs precision data and the direct search bounds
are usually checked by
the public codes of \textsc{HiggsSignals}~\cite{Bechtle:2020uwn} and \textsc{HiggsBounds}~\cite{Bechtle:2020pkv}, respectively.
However,
even the most recent version \textsc{HiggsBounds}-v5.10.2 
misses three important processes for a light fermiophobic Higgs boson,
multi-photon signals measured by the DELPHI~\cite{DELPHI:2003hpv},
CDF~\cite{CDF:2016ybe}, and CMS collaborations~\cite{CMS:2021bvh}.
We need to include the three processes in the random scanning.
Based on the finally allowed parameter points,
we will perform for the first time the RGE analysis to obtain the cutoff scale of every viable parameter point.
In addition, we will study
the decays and productions of the BSM Higgs bosons to suggest
$4\gm+VV'$ as the golden discovery channel at the HL-LHC,
which complements the studies of
$4\gm+V$ in the literature~\cite{Akeroyd:2003bt,Akeroyd:2005pr,Akeroyd:2003xi,Delgado:2016arn,Arhrib:2017wmo}.
In addition to presenting the cross section times branching ratio as well as the selection efficiencies
at the detector level,
we will also present the correlation between the cutoff scale and the signal rates.
These are our new contributions.

The paper is organized in the following way. 
  In Sec.~\ref{sec:review}, we give a brief review of the fermiophobic type-I in the 2HDM.
  In Sec.~\ref{sec:scan}, we do the parameter scanning by imposing the theoretical and experimental
  constraints, including the RGE analysis.
    Section \ref{sec:LHC} deals with the LHC phenomenology of the BSM Higgs bosons. 
   Finally we conclude in Sec.~\ref{sec:conclusions}.  

\section{Fermiophobic $h$ in type-I 2HDM}
\label{sec:review}

In the 2HDM, there exist two $SU(2)_L$ complex scalar doublet fields with hypercharge $Y=+1$~\cite{Branco:2011iw}:
\bea
\label{eq:phi:fields}
\Phi_i = \left( \begin{array}{c} w_i^+ \\[3pt]
\dfrac{v_i +  \rho_i + i \eta_i }{ \sqrt{2}}
\end{array} \right), \quad (i=1,2)
\eea
where $v_{1}$ and $v_2$ are the vacuum expectation values of $\Phi_{1}$ and $\Phi_2$ respectively.
The electroweak symmetry is spontaneously broken by the nonzero vacuum expectation value 
of $v =\sqrt{v_1^2+v_2^2}=246\gev $.
The ratio of $v_2$ to $v_1$ is defined by $\tan\beta = v_2/v_1$ ($\bt \in [0,\pi/2]$).
For notational simplicity, we take $s_x =\sin x$, $c_x = \cos x$, and $t_x = \tan x$ in what follows.
A discrete $Z_2$ symmetry is introduced to prevent the tree level flavor-changing neutral currents 
(FCNC)~\cite{Glashow:1976nt,Paschos:1976ay},
under which $\Phi_1 \to \Phi_1$ and $\Phi_2 \to -\Phi_2$.
Then the scalar potential with softly broken $Z_2$ symmetry and \textit{CP}-invariance is
\bea
\label{eq:VH}
V_\Phi = && m^2 _{11} \Phi^\dagger_1 \Phi_1 + m^2 _{22} \Phi^\dagger _2 \Phi_2
-m^2 _{12} ( \Phi^\dagger_1 \Phi_2 + \hc) \\ \nn
&& + \frac{1}{2}\lambda_1 (\Phi^\dagger _1 \Phi_1)^2
+ \frac{1}{2}\lambda_2 (\Phi^\dagger _2 \Phi_2 )^2
+ \lambda_3 (\Phi^\dagger _1 \Phi_1) (\Phi^\dagger _2 \Phi_2)
+ \lambda_4 (\Phi^\dagger_1 \Phi_2 ) (\Phi^\dagger _2 \Phi_1) \\ \nn
&& + \frac{1}{2} \lambda_5
\left[
(\Phi^\dagger _1 \Phi_2 )^2 +  \hc
\right].
\eea

Five physical Higgs bosons exist, the light \textit{CP}-even scalar $h$,
the heavy \textit{CP}-even scalar $H$, the \textit{CP}-odd pseudoscalar $A$,
and a pair of charged Higgs bosons $H^\pm$.
The mass eigenstates are related with the weak eigenstates in Eq.~(\ref{eq:phi:fields})
via two mixing angles of $\al$ and $\bt$,
of which the relations are referred to Ref.~\cite{Song:2019aav}.
The SM Higgs boson is a linear commination of $h$ and $H$, given by
\bea
\label{eq:hsm}
\hsm = \sba h + \cba H.
\eea
The observed SM-like Higgs boson at a mass of $125\gev$ can be either $h$ or $H$.
In this work, we concentrate on the inverted scenario where $\mhh=125\gev$.
Then the Higgs coupling modifiers to $V$ ($V=\wpm,Z$) are 
\bea
\label{eq:kpV:xiV}
\kp^H_V =\cba,
\quad
\xi^h_V = \sba.
\eea
Note that the SM-like Higgs boson demands $|\sba | \ll 1$.
For the sign of $\cba$, we adopt the scheme with $\cba>0$ as in the public codes such as \textsc{2HDMC}~\cite{Eriksson:2009ws},
\textsc{HiggsSignals}~\cite{Bechtle:2020uwn}, and \textsc{HiggsBounds}~\cite{Bechtle:2020pkv}.

We parameterize the Yukawa interactions of the SM fermions as
\bea
\label{eq:Lg:Yukawa}
\lg^{\rm Yuk} &=&
- \sum_f 
\lf 
\frac{m_f}{v} \xi^h_f \bar{f} f h + \frac{m_f}{v} \kp_f^H \bar{f} f H
-i \frac{m_f}{v} \xi_f^A \bar{f} \gm_5 f A
\ri
\\ \nn &&
- 
\left\{
\dfrac{\sqrt{2}}{v } \overline{t}
\left(m_t \xi^A_t \text{P}_- +  m_b \xi^A_b \text{P}_+ \right)b  H^+
+\dfrac{\sqt m_\tau}{v}\xi^A_\tau \,\overline{\nu}_\tau P_+ \tau H^+
+\hc
\right\},
\eea
where $f=t,b,\tau$ and $P_\pm = (1 \pm \gm^5)/2$.
In type-I, the normalized Yukawa couplings are
\begin{align}
\label{eq:Yukawa:couplings}
\xi^h_f&=\frac{\ca}{\sb} ,
\quad
\kp^H_f = \frac{\sa}{\sb} ,
\quad
\xi^A_t = -\xi^A_b = -\xi^A_\tau = \frac{1}{\tb}.
\end{align}
If $\al=\pi/2$, $\xi^h_f=0$ so that $h$ becomes fermiophobic.
We assume that the fermiophobic condition is preserved at loop level
by a suitable renormalization condition~\cite{Barroso:1999bf,Brucher:1999tx}.
To emphasize the fermiophobic nature,
we take the notation of $\hf$ for the light \textit{CP}-even Higgs boson with vanishing Yukawa couplings to the SM
fermions.
Our model is summarized as
\bea
\label{eq:setup}
\mhh=125\gev , \quad \al = \frac{\pi}{2}.
\eea
Since $\tb = -{\cba}/{\sba}$ if $\al=\pi/2$,
the condition of $|\sba| \ll 1$ corresponds to $\tb \gg 1$.

\section{Scanning strategies and model validity}
\label{sec:scan}

We scan the parameter space by imposing all the theoretical and experimental constraints.
To efficiently satisfy the oblique parameters of $S$, $T$, and $U$,
we make the assumption of $\ma=\mch\equiv \mach$.
Then there are four model parameters, $\mhf$, $\mach$, $\tb$, and $m_{12}^2$.
For the fixed $\mhf$ of
\bea
\label{eq:mhf}
\mhf=20,~30,~40,~60,~96\gev,
\eea
we randomly generate the other parameters over the ranges of
\bea
\label{eq:scan:range}
&& 
\mach \in \left[ 80,\, 900 \right] \gev,  \\ \nn
&& \tb  \in \left[ 1,\,100 \right], \qquad
m_{12}^2 \in \left[ 0, 15000 \right] \gev^2.
\eea

Now we cumulatively impose the following constraints:
\begin{description}
\item[Step-A.] \textbf{Theoretical constraints with low energy data:}
\renewcommand\labelenumi{(\theenumi)}
	\ben
	\item We require the Higgs potential being bounded from below~\cite{Ivanov:2006yq},
the unitarity of scalar-scalar scatterings~\cite{Branco:2011iw,Arhrib:2000is},
the perturbativity of Higgs quartic couplings~\cite{Chang:2015goa}, and the stability of the \textit{CP}-conserving vacuum~\cite{Ivanov:2008cxa,Barroso:2012mj,Barroso:2013awa}.
The public code \textsc{2HDMC}~\cite{Eriksson:2009ws} is used.
	\item We demand the Peskin-Takeuchi electroweak oblique parameters $S$, $T$, and $U$~\cite{Peskin:1991sw}
	to satisfy the current best-fit results~\cite{Zyla:2020zbs} at 95\% C.L.:
	\bea
	\label{eq:STU:PDG}
	S &=& -0.01 \pm 0.10,
	\quad
	T = 0.03 \pm 0.12, \quad
	U=0.02 \pm 0.11, 
	\\ \nn
	 \rho_{ST} &=& 0.92, \quad \rho_{SU}=-0.80,\quad \rho_{TU}=-0.93,
	\eea
	where $\rho_{ij}$ is the correlation matrix.
	We use the 2HDM calculations of $S$, $T$, and $U$ in Refs.~\cite{He:2001tp,Grimus:2008nb,Zyla:2020zbs}.
	\item We require that the most recent constraints from flavor physics 
	 be satisfied at 95\% C.L.~\cite{Arbey:2017gmh,Misiak:2017bgg}.
	An important constraint is from $b\to s \gm$, which excludes the region with small $\tb$ 
	and the light charged Higgs boson: for example,
$\tan \beta > 2.6$ for $M_{H^+} = 140\gev$~\cite{Arbey:2017gmh}. 
	\item We demand that the model  be valid at least up to $1\tev$,
	i.e., the cutoff scale $\lmc$ be larger than 1 TeV.
	$\lmc$ is obtained through the following procedures:
 		\bit
		\item For each parameter point,
		we perform the RGE running at one loop level,
		starting from the top quark pole mass scale, $m_{t}^{\rm pole} = 173.4\gev$.
		The boundary conditions at $m_{t}^{\rm pole}$ are referred to Ref.~\cite{Oredsson:2018yho}.
		We use the open code \textsc{2HDME}~\cite{Oredsson:2018vio}.
		\item At the next high energy scale,
		we check three conditions---unitarity, perturbativity, and vacuum stability.
		If any condition is broken,
		we stop the running and record the energy scale as the cutoff scale $\lmc$.		
		\eit
	\een
\item[Step-B.] \textbf{High energy collider data:}
\renewcommand\labelenumi{(\theenumi)}
	\ben
	\item We check whether each parameter point satisfies the Higgs precision data.
	We use the public code \textsc{HiggsSignals}-v2.6.2~\cite{Bechtle:2020uwn},
	which gives the $\chi^2$ value for 111 Higgs
observables~\cite{Aaboud:2018gay,Aaboud:2018jqu,Aaboud:2018pen,Aad:2020mkp,Sirunyan:2018mvw,Sirunyan:2018hbu,CMS:2019chr,CMS:2019kqw}.
	We demand that the $p$-value should be larger than 0.05.
	\item The parameter point should not conflict with the null results of
	the direct searches at the LEP, Tevatron, and LHC.
Using the open code \textsc{HiggsBounds}-v5.10.2~\cite{Bechtle:2020pkv},
we exclude a parameter point
if the predicted cross section is larger than 95\% C.L. upper bound on the observed cross section.
	\item Three important processes for a light fermiophobic Higgs boson are missing in \textsc{HiggsBounds},
	the DELPHI measurement of $\rr Z/\rr\bb$ states~\cite{DELPHI:2003hpv},
	the CDF measurement of $4 \gm \wpm$~\cite{CDF:2016ybe},
	and the CMS search for $4\gm$~\cite{CMS:2021bvh}.
	We include the constraints from these processes.
	\een
\end{description}

\begin{table*}[!t]
\setlength\tabcolsep{10pt}
\centering
{\footnotesize\renewcommand{\arraystretch}{1.1} 
\begin{tabular}{|c||c|c|c|c|c|}
\toprule
& \multicolumn{5}{c|}{Survival probabilities}  \\ \hline
$\mhf$ [GeV] & 20 & 30 & 40 & 60 & 96\\
\hline
\hline
Step-B(2) & 1.10\% & 0.27\% & 0.13 \% & 0.026 \%  & 25.7\%\\
\hline
Step-B(3) & 0.207\% & 0.048\% & 0.011\% & 0.000\% & 25.7\%\\
\bottomrule
\end{tabular}
}
\caption{Survival probabilities at Step-B(2) and Step-B(3)
for $\mhf=20,30,40,60,96\gev$.
The reference is the parameter point that pass Step-A.
}
\label{tab:survival}
\end{table*}

For each $\mhf$ in \eq{eq:mhf}, we first obtained $1.2 \times 10^6$ parameter points that pass Step-A,
over which we imposed the constraints at Step-B.
The Higgs precision data and the direct search bounds severely constrain the model.
The three processes in Step-B(3) additionally remove the parameter points.
To quantify the curtailment of the parameter space,
we separately show the survival probabilities at Step-B(2) and 
Step-B(3) in Table~\ref{tab:survival}, with respect to the parameter points after Step-A.
We first focus on Step-B(2).
For $\mhf \lsim \mhsm/2$, 
the fermiophobic type-I barely survives with the probability below $\mco(1)\%$.
The killer processes are the ATLAS measurement of $\hsm \to \hf\hf\to 4\gm$~\cite{ATLAS:2015rsn}
and the LEP measurement of $\ee\to \hf Z\to \rr+Z$~\cite{Rosca:2002me}.
For $\mhf=96\gev$, on the other hand, a large portion of the parameter space remains valid
with the survival probability of about 25\% at Step-B(2).
The smoking-gun processes are 
 the ATLAS measurements of $t\to \ch b \to \tau\nu+b$~\cite{ATLAS:2018gfm}
and $pp\to \hf/W\hf/Z\hf/tt\hf \to \rr+X$~\cite{ATLAS:2018xad}.

Let us discuss the three processes at Step-B(3).
The first is the DELPHI search for a fermiophobic Higgs boson in the multi-photon states~\cite{DELPHI:2003hpv},
which consists of $\ee\to \hf(\to \rr) Z$
and $\ee\to \hf(\to \rr ) A(\to \bb/\hf Z) $.
For $\mhf=30\gev$, the DELPHI results exclude $\ma \leq 148\gev$,
which reduces the survival probability at Step-B(2) by about 80\%.
For $\mhf=96\gev$, however,
the DELPHI results do not affect because of the heavy mass.
The second is the CDF measurement of $p\bar{p}\to \ch(\to  \wpm \hf) \hf \to 4 \gm+X$~\cite{CDF:2016ybe}.
Based on the assumptions of $\br(\hf \to \rr)=1$ and $\ma=350\gev$,
the CDF collaboration presented the excluded region of $(\mch,\mhf)$:
for $\mhf=30\gev$, $\mch \leq 160\gev$ is prohibited.
But the case of $\mhf=96\gev$ is not affected by the CDF measurement
since $\br(\hf\to \rr) \lsim 0.25$ as shall be shown below.
The third is the CMS search for the exotic decay of the Higgs boson into four-photon
final states in the mass range of $\mhf \in [10,60]\gev$~\cite{CMS:2021bvh}:
for $\mhf=30\gev$, $\sg\times \br \leq 0.9\fb$.
An extraordinary case is for $\mhf=60\gev$.
Out of $1.2\times 10^6$ points after Step-A,
only about 300 points survive Step-B(2) and no parameter point remains after Step-B(3).
The smoking-gun processes at Step-B(2)
are the exotic Higgs decay of $\hsm \to \hf\hf \to 4\gm$
and the LHC process of $ gg\to A \to Z^{(*)}(\to \elll)\hsm(\to \bb)$~\cite{ATLAS:2020pgp},
which permits $\mch \in [80,\,120]\gev$.
Then the DELPHI and CDF results exclude the light $\mach$ region.

% Fig 1
\begin{figure}[t] \centering
\begin{center}
\includegraphics[width=0.56\textwidth]{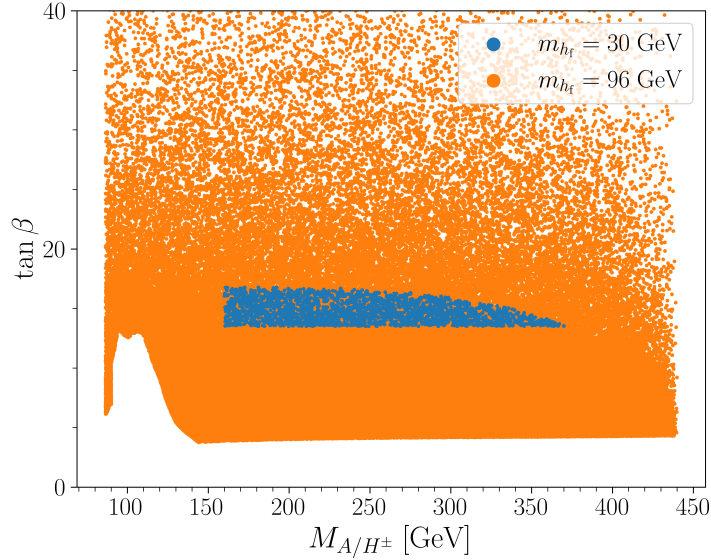}
\end{center}
\vspace{-0.7cm}
\caption{\label{fig-tb-MA}
$\tan\beta$ versus $\mach$ for $\mhf=30\gev$ (blue points) and $\mhf=96\gev$ (orange points).
}
\end{figure}

We investigate the characteristics of the finally allowed parameter space.
Figure~\ref{fig-tb-MA} shows $\tb$ versus $\mach$ 
for $\mhf=30\gev$ (blue points) and $\mhf=96\gev$ (orange points).
Both cases have the upper bounds on the masses of $A$ and $\ch$: 
$M_{A/\ch}\lsim 360\gev$ for $\mhf=30\gev$ and $M_{A/\ch}\lsim 440\gev$ for $\mhf=96\gev$.
The lower bound on $M_{A/\ch}$ also exists,
$\mach \gsim 160\gev$ for $\mhf=30\gev$ and $\mach \gsim 90\gev$ for $\mhf=96\gev$.
The larger lower bound on $\mach$ for $\mhf=30\gev$
is from the constraints at Step-B(3).
In addition, we have substantially large $\tb$ for $\mhf=30\gev$, above about 13. 
For $\mhf=96\gev$, however, $\tb$ can be as small as about 3. 
The relation of $\tb = - \cba/\sba$ in the fermiophobic type-I
translates $\tb$ into $\sba$.
We have  $|\sba| \lsim 0.08$ for $\mhf=30\gev$ and $|\sba| \lsim 0.27$ for $\mhf=96\gev$.
Considerable deviation from the Higgs alignment can occur when $\mhf$ is heavier than half the observed Higgs boson mass.

% Fig 2
\begin{figure}[t] \centering
\begin{center}
\includegraphics[width=0.6\textwidth]{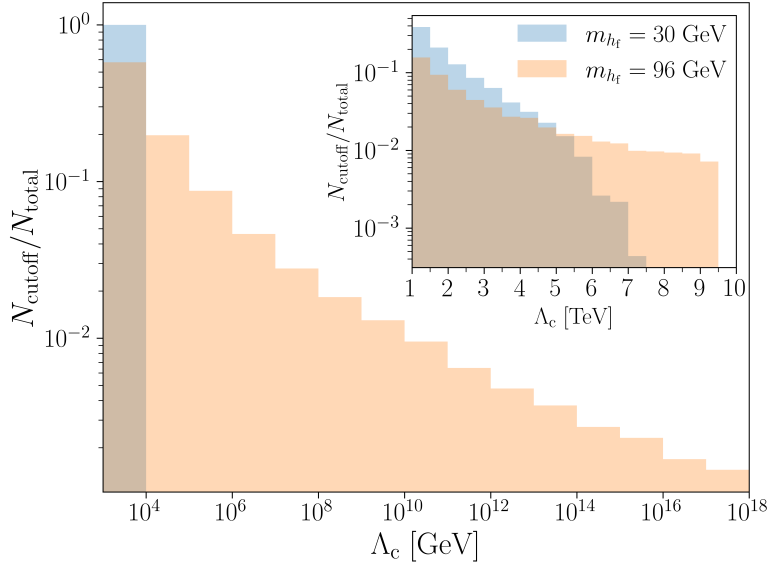}
\end{center}
\vspace{-0.7cm}
\caption{\label{fig-RGE}
Distributions of the cutoff scales of the finally allowed parameter points 
for $\mhf=30,~96\gev$.
}
\end{figure}

Although the fermiophobic type-I is shown to satisfy all the theoretical and experimental constraints, 
this validity is only at the electroweak scale.
Then to what energy scale is this model valid?
Focusing on two cases with $\mhf=30,~96\gev$,
we calculate the cutoff scale for each parameter point after Step-B(3).
Figure~\ref{fig-RGE} shows the distribution of $\lmc$.
The $y$-axis represents the ratio $N_{\rm cutoff}/N_{\rm total}$,
where  $N_{\rm cutoff}$ is the number of the parameter points with the cutoff scale $\lmc$ and 
$N_{\rm total}$ is the total number of the finally allowed parameter points.
It is clearly seen that the case of $\mhf=30\gev$ has quite low cutoff scales.
A very light fermiophobic Higgs boson is not only difficult to satisfy the current data
(see the small survival probabilities in Table~\ref{tab:survival}) but also unable to maintain stability above $10\tev$.
It demands an extension at the energy scale not far from the LHC reach.
If $\mhf=96\gev$,
however, 
the scalar potential can be stable all the way up to the Planck scale.
The parameter ranges with $\lmc>10^{18}\gev$ are
\bea
\label{eq:Planck}
\mhf=96\gev \hbox{ with }\lmc>10^{18}\gev:&& \mach \in   [94.9, \, 149.7] \gev,
\\ \nn
&&    m_{12}^2 \in  [ 97.8,\, 2071.7] \gev^2,
\\ \nn &&
\sba \in [-0.22, \,-0.01],\quad 
\tb \in [4.34,\, 94.3],
\eea
which prefer intermediate masses for $A$ and $\ch$.

Finally, we make brief comments on the recent measurement 
of the $W$ boson mass by the CDF collaboration,
$m_W^{\rm CDF} = 80.4335 \pm 0.0094\gev$~\cite{CDF:2022hxs}.
If we take $m_W^{\rm CDF} $ as a face value,
it shows $7\sg$ deviation from the SM prediction, 
which yields different oblique parameters as
$S_{\rm CDF}=0.15\pm 0.08$ and $T_{\rm CDF}=0.27\pm 0.06$ with $U=0$~\cite{Lu:2022bgw}. 
Although the implications of the anomaly in the framework of the 2HDM have been extensively studied~\cite{Fan:2022dck,Zhu:2022tpr,Lu:2022bgw,Zhu:2022scj,Song:2022xts,Bahl:2022xzi,Heo:2022dey,Babu:2022pdn,Biekotter:2022abc,Ahn:2022xeq,Han:2022juu,Arcadi:2022dmt,Ghorbani:2022vtv,Broggio:2014mna,Lee:2022gyf,Kim:2022hvh},
the fermiophobic type-I has not been considered in the literature.
So we rescanned the parameter space with $S_{\rm CDF}$ and $T_{\rm CDF}$.
With the condition of $\dma \lf \equiv \ma-\mch \ri=0$,
no parameter point survives even Step-A(2).
If the mass degeneracy is relaxed to $\dma =  -20\gev$,
many parameter points pass Step-A(3), but no point satisfies the condition of $\lmc>1\tev$ at Step-A(4).
If $\dma =-30\gev$, 
a large portion of the parameter space passes the final step.
In summary, the CDF $W$ boson mass measurement demands a sizable (but not large) mass gap
between $A$ and $\ch$.

\section{LHC phenomenology}
\label{sec:LHC}

Now that the finally allowed parameter space is significantly limited,
the observation of the fermiophobic type-I at the LHC 
is more predictive than ever.
We first discuss the decays of $\hf$,
which are only into electroweak bosons:
the loop-induced decay of $\hf\to gg$ also vanishes because it occurs through quarks.
The off-shell decays of $\hf$ into $VV^*$
is feasible due to the deviation from the Higgs alignment, especially in the case of $\mhf=96\gev$.
The radiative decay of $\hf$ into $\rr$ gets the contribution
from $W^\pm$ and $\ch$ in the loop.
The coupling for the $\hf$-$H^+$-$H^-$ vertex is
\bea
\lmh_{hH^+ H^-} = \frac{1}{v^2}
\left[
(2 \mch^2 - \mhf^2) \cb  + \frac{\mhf^2 - M^2}{ \cb} 
\right]
\simeq   \frac{\tb}{v^2}
 (\mhf^2 - M^2),
\eea
where $M^2 = m_{12}^2/(\sb\cb)$ and
$\lmh_{hH^+ H^-} $ is defined by $\lg_{\rm tri} \supset v \lmh_{hH^+ H^-} \hf H^+ H^-$.
Since the finally allowed parameter points prefer large $\tb$ and intermediate  $\mch$,
the decay mode of $\hf \to \rr$ will be important.

The branching ratios of the BSM Higgs bosons crucially depend on the mass of $\hf$.
Let us first consider the case of $\mhf=30\gev$.
Very light $\hf$ exclusively decays into $ \rr$ with $\br(\hf \to \rr)\simeq 100\%$.
The other BSM Higgs bosons, $A$ and $\ch$,
also have a single dominant decay mode,
$A \to Z\hf$ and $\ch\to \wpm \hf$, which have almost 100\% branching ratios.
The main reason is the kinematic gain from the light $\mhf$ and the heavy enough $\mach$
as shown in Fig.~\ref{fig-tb-MA}.

% Fig 3
\begin{figure}[t!] \centering
\begin{center}
\includegraphics[width=\textwidth]{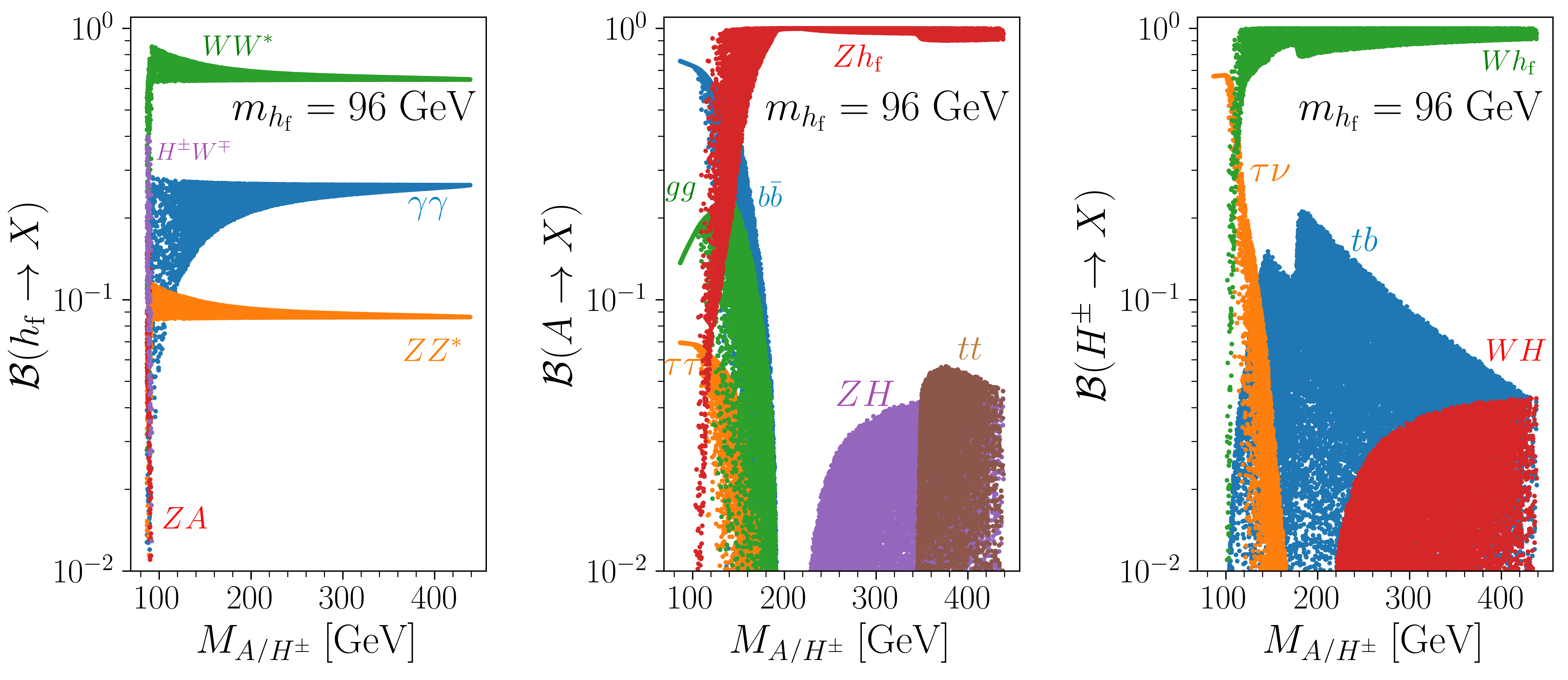}
\end{center}
\vspace{-0.7cm}
\caption{\label{fig-BR}
Branching ratios of $\hf$ (left panel), $A$ (middle panel), and $\ch$ (right panel) 
for $\mhf=96\gev$.
}
\end{figure}

In the case of $\mhf=96\gev$, 
the decay patterns of the BSM Higgs bosons are more diverse.
Figure \ref{fig-BR} shows the branching ratios of $\hf$ (left panel), $A$ (middle panel), and  $\ch$ (right panel) 
as a function of $\mach$,
over the finally allowed parameter points.
The leading decay mode of $\hf$ is into $WW^*$ with the branching ratio of about 70\%.
We ascribe it to the sizable deviation from the Higgs alignment and the tree-level decay.
The diphoton mode is next-to-leading with $\br(\hf\to \rr)\lsim 28\%$.
The third dominant decay mode is $\hf \to ZZ^*$ with the branching ratio of about 10\%.
The pseudoscalar $A$ has the most diverse decay modes among the BSM Higgs bosons.
For $\ma \lsim 120\gev$,
$A\to \bb$ is the leading decay mode, followed by $A\to gg$ and $A\to \ttau$.
For  $\ma \gsim 150\gev$,
the bosonic mode of $A\to Z \hf$ becomes dominant.
Above the kinematic threshold,
the decays of $A\to Z H$ and $A\to \ttop$ can be substantial
with a few percent branching ratios.
The charged Higgs boson mainly decays into $\wpm \hf$~\cite{Cheung:2022ndq}
except for small mass window of $\mch \in [80, \,110]\gev$ where $\ch\to\tau\nu$ is dominant.
Above the kinematic threshold, 
the branching ratios of $\ch\to t b$ and $\ch \to \wpm H$ can be of the order of 1\%.

Brief comments on the possibility of $\hf$ as a long-lived particle are in order here.
If the total decay width of $\hf$ 
is small enough to make $\hf$ long-lived,
we see a signal of a photon pair from the secondary vertex displaced from the primary vertex.
If a particle has  $\Gm_\tot= 10^{-11}\gev$, for example,
the decay length $d \simeq  \bt\gm \times 10^{-6}{\rm m}$
where $\gm = E/m$ and $\beta=v/c$.
Since photons do not leave any track in the inner tracker,
we have to deduce the secondary vertex from the photon tracks in the ECAL.
In Ref.~\cite{Banerjee:2019ktv},
it was shown that we need the decay length larger than 0.1 cm to probe the signal,
which require $\Gm_\tot <10^{-13}\gev$ even with $\gm=10$.
We find that the minimum of $\Gm_\tot^{\hf}$ is $1.3\times 10^{-12}\gev$ for $\mhf=20\gev$,
$3.9\times 10^{-12}\gev$ for $\mhf=30\gev$,
and
$1.9\times 10^{-9}\gev$ for $\mhf=96\gev$.
It is difficult to see the displaced diphoton signals of a long-lived $\hf$ at the LHC in the fermiophobic type-I.

Considering the branching ratios of the BSM Higgs bosons,
we study the following multi-photon states at the LHC:
\begin{eqnarray}
\label{eq:4rV}
4\gm+Z: && p p \to   A (\to Z\hf) \hf \to 4\gm +Z,  \\
\nn
	&& pp \to H (\to \hf \hf) Z \to 4\gm +Z,\\[5pt]
\nn
4\gm +\wpm : && pp \to \ch(\to \wpm \hf) \hf \to 4\gm +\wpm,\\   \nonumber
&& p p \to  H (\to \hf \hf) \wpm \to 4\gm +\wpm,\\[5pt] 
\label{eq:4rVV'}
4\gm +\wpm Z: && p p \to  \ch (\to \wpm \hf) A(\to Z \hf) \to 4\gm + \wpm Z,
\\[5pt]  \nn
4\gm +W^+ W^- : && p p \to  H^+ (\to W^+ \hf) H^- (\to W^- \hf) \to 4\gm +W^+W^-.
\end{eqnarray}
Note that the Higgs alignment favors the productions of $\qq\to Z^* \to HA$  and
$q\bar{q}' \to W^* \to \ch A$.

\begin{figure}[t] \centering
\begin{center}
\includegraphics[width=\textwidth]{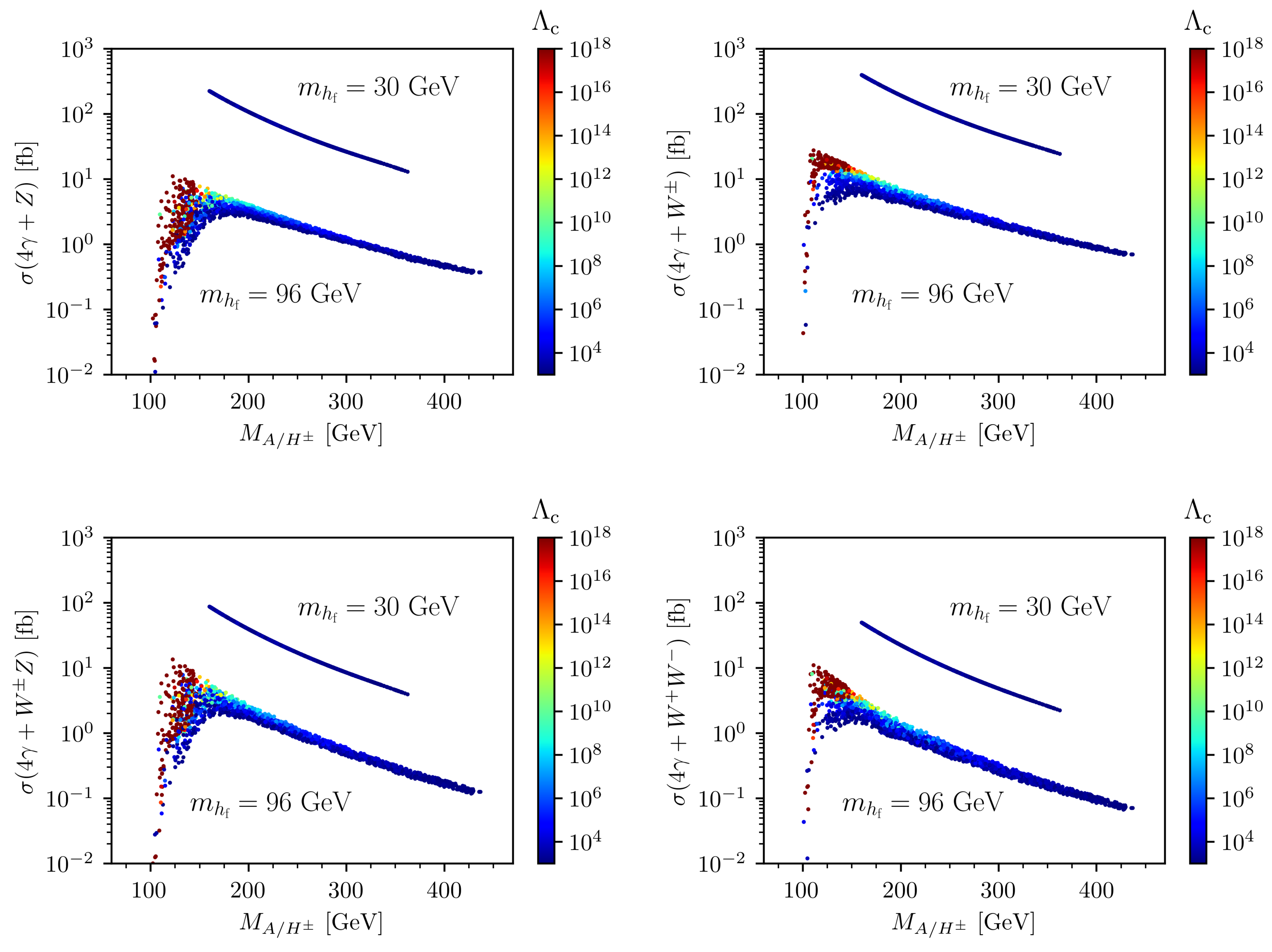}
\end{center}
\vspace{-0.7cm}
\caption{\label{fig:cross_sections}
Cross-section times branching ratio of multi-photon states as a function of $\mach$:
$4\gm+Z$ (upper-left panel), $4\gm+\wpm$ (upper-right panel), 
$4\gm+ \wpm Z$ (lower-left panel), 
and $4\gm+W^+ W^-$ (lower-right panel) for $\mhf=30\gev$ and $\mhf=96\gev$.
The color code denotes the cutoff scale $\lmc$.
}
\end{figure}

In Fig.~\ref{fig:cross_sections}, we present as a function of $\mach$
the production cross section times branching ratio
for $4\gm+Z$ (upper-left panel), $4\gm+\wpm$ (upper-right panel), 
$4\gm+\wpm Z $ (lower-left panel), 
and $4\gm+W^+ W^-$ (lower-right panel) over the finally allowed parameter points.
We take two cases of $\mhf=30\gev$ and $\mhf=96\gev$.
The color code denotes the cutoff scale $\lmc$.
To calculate the parton-level production cross sections, 
we first use \textsc{FeynRules}~\cite{Alloul:2013bka} 
to obtain the Universal FeynRules Output (UFO)~\cite{Degrande:2011ua} for the  fermiophobic type-I.
Interfering the UFO file with \textsc{MadGraph5-aMC@NLO}~\cite{Alwall:2011uj},
we compute the cross-sections of $pp \to A\hf/HZ/\ch \hf/H \wpm/\ch A/H^+H^-$ at 14 TeV LHC using \textsc{NNPDF31\_lo\_as\_0118} parton distribution function set~ \cite{NNPDF:2017mvq}. 
%We include the gluon fusion production, although the loop-induced production is subleading.
The two-body cross-sections are multiplied by relevant branching ratios of $\hf$, $A$, and $H^\pm$
obtained from the \textsc{2HDMC}~\cite{Eriksson:2009ws}.
%For the cross check of the results in \fig{fig:cross_sections}, 
%we first include the effective vertices of the Higgs bosons to a gluon pair and a photon pair
%at the level of \textsc{FeynRules} to obtain the UFO file.
%Having the BSM Higgs bosons decay at \texttt{MadGraph5-aMC@NLO},
%we calculate the parton-level cross sections of $4\gm+V$ and $4\gm+VV'$,
%and find the consistency.

The first noteworthy result in Fig.~\ref{fig:cross_sections} is 
that the signal rate of $\mhf=30\gev$ is about ten times larger than that of $\mhf=96\gev$ for all the processes.
This feature is attributed primarily to the kinematic advantage of very light $\mhf$.
Unexpected is the result that the signal rate of $4\gm +VV'$ is almost compatible with
that of $4\gm +V$, even though $4\gm +VV'$ has one more particle in the final state.
It is because both $4\gm+V$ and $4\gm +VV'$ are basically $2\to 2$ processes with large branching ratios: 
see Eqs.~(\ref{eq:4rV})
and (\ref{eq:4rVV'}).
Since $4\gm +VV'$ has one more tagging particle, which is hard for the SM backgrounds to mimic,
we suggest $4\gm +VV'$ as the golden discovery channel for the light fermiophobic Higgs boson.

Now, let us discuss the correlation of the signal rate to cutoff scale.
For the case of $\mhf=30\gev$, which has a low cutoff scale below $7\tev$,
there is no point in discussing the correlation.
In the case of $\mhf=96\gev$, however,
the cutoff scale is widely distributed, from $1\tev$ to the Planck scale.
So, the correlation has a profound implication on the LHC phenomenology.
The color code of $\lmc$ in Fig.~\ref{fig:cross_sections} clearly shows that 
the maximum signal rate for all the multi-photon final states
happens when the model has $\lmc\sim 10^{18}\gev$.

A critical question is whether we can discover the fermiophobic $\hf$
via the multi-photon signals at the HL-LHC.
We find that the SM backgrounds are almost negligible.
Let us consider the process of $4\gm+\wpm$ as an example.
The irreducible background,
demanding four photons and one lepton with $p_T^\gm >10\gev$, 
$p_T^\ell >20\gev$, $|\eta^{\gm,\ell}|<2.5$, and 
the isolation cuts of $\Dt R >0.4$ where $\Dt R = \sqrt{(\Dt \eta)^2 + (\Dt \phi)^2}$
has the cross section about $1\ab$~\cite{Arhrib:2017wmo}.
If we require two gauge bosons associated with four photons,
the irreducible backgrounds are totally negligible.
The reducible backgrounds are from QCD jets accompanying one or two gauge bosons,
with the QCD jet misidentified as a photon.
But the mistagging probability is very small as $P_{j \to \gm} \simeq 10^{-3}$~\cite{Wang:2021pxc}.
The final states with four photons require the multiplication of $P_{j \to \gm} $ four times,
which makes the reducible backgrounds also ignorable.
Therefore, we conclude that the $4\gm +VV'$ states have an almost background-free environment.

The final question is about the selection efficiencies of $4\gm+VV'$.
No matter how negligible the backgrounds are,
small cross sections raise concerns about whether we can observe enough signal events
especially when the selection efficiencies are very small.
Full state-of-the-art simulations, customized for $4\gm+W^+W^-$ and $4\gm+\wpm Z$,
are beyond the scope of this letter.
Here we present the detector-level selection efficiency
only for four photons.
Through a fast detector simulation of the signal
using the \textsc{Delphes} version 3.4.2~\cite{deFavereau:2013fsa},
we calculate $\es_{\rm d} = \sg({\rm cuts})/\sg({\rm no\;cuts})$.
We demand $p_T^\gm > 10\gev$~\cite{Arhrib:2017wmo}, $|\eta^\gm|<2.5$,
and $\Dt R(\gm_i,\gm_j)>0.4$.
%but exclude the region of $1.37<|\eta|<1.52$ 
%where the barrel and endcap of the EM calorimeter are overlapped~\cite{ATLAS:2018fzd}.
We take the following four benchmark points:
\bea
\label{eq:BP}
\hbox{BP1: } && \mhf=30\gev, \quad \mach = 355.5\gev, \quad \tb=13.6,\quad m_{12}^2 = 1.07\gev^2,
\\ \nn
\hbox{BP2: } && \mhf=30\gev, \quad\mach =164.0 \gev,  \quad \tb=14.6,\quad m_{12}^2 = 0.97\gev^2,
\\ \nn
\hbox{BP3: } && \mhf=96\gev, \quad \mach =428.3 \gev, \quad \tb=4.8,\quad m_{12}^2 = 1663.0\gev^2,
\\ \nn
\hbox{BP4: } && \mhf=96\gev, \quad \mach =150.1 \gev, \quad \tb=26.6,\quad m_{12}^2 = 344.7\gev^2,
\eea
where BP1 and BP3 yield small signal rate of $4\gm+VV'$ for $\mhf=30\gev$ and $\mhf=96\gev$ respectively
while BP2 and BP4 yield large signal rate.

\begin{table*}[!t]
\setlength\tabcolsep{10pt}
\centering
{\footnotesize\renewcommand{\arraystretch}{1.1} 
\begin{tabular}{|cc||c|c||c|c|}
\toprule
 \multicolumn{6}{|c|}{selection efficiency $\es_{\rm d} = \sg({\rm cuts})/\sg({\rm no\;cuts})$}  \\ \hline
& &  \multicolumn{2}{c|}{$4\gm+WW$} &  \multicolumn{2}{c|}{$4\gm+WZ$} \\
\hline
\hline
 & $(\mhf,\mach)$\;[GeV] & $n_\gm=3$ & $n_\gm=4$ & $n_\gm=3$ & $n_\gm=4$ \\
 \hline
 BP1 & $(30\gev,\, 355.5)$ & 13.3\% & ~7.2\%& 13.1\%&  ~7.2\%\\
 BP2 & $(30\gev,\, 164.0)$ &  30.7\% & 15.1\% & 31.3\% & 14.3\% \\
 BP3 &  $(96\gev,\, 428.3)$ & 31.1\%  & 39.9\% & 31.6\% & 40.1\% \\
 BP4 &  $(96\gev,\, 150.1)$ &  36.3\% & 27.3\% & 36.6\% & 27.9\% \\
 \bottomrule
\end{tabular}
}
\caption{Selection efficiencies for four benchmark points 
for the $4\gm+W^+W^-$ and $4\gm+\wpm Z$ final states,
including  the off-shell $V$.
Here $n_\gm$ is the number of photons that pass the cuts. 
The cuts are $p_T^\gm > 10\gev$, $|\eta^\gm|<2.5$, and $\Dt R(\gm_i,\gm_j)>0.4$ at the detector level.
}
\label{tab:efficiency}
\end{table*}

Table~\ref{tab:efficiency} shows the selection efficiency $\es_{\rm d}$
for the $4\gm+W^+W^-$ and $4\gm+Z\wpm$ final states, focusing on four benchmark points in \eq{eq:BP}.
Here we include the off-shell $W$ and $Z$ bosons.
Since some photons decayed from the light $\hf$ can fail to pass the cuts at the detector level,
we separately present the selection efficiencies for $n_\gm=3$ and $n_\gm=4$,
where $n_\gm$ is the number of photons that pass the cuts.
In the case of $\mhf=30\gev$,
the selection efficiency with $n_\gm=3$ is larger than that with $n_\gm=4$.
The light $\mhf$ tends to yield soft photons so that it is more probable for one of four photons to escape the detection.
This tendency is stronger for BP2 where the smaller mass difference between $\mhf$ and $\mach$
generates softer photons.
If we relax the photon identification into $n_\gm \geq 3$,
therefore,
the selection efficiency significantly increases:
even the benchmark BP1 has the selection rate of about 20\%,
which can have  a few thousand events with the total integrated luminosity of $3\ab^{-1}$.
In the case of $\mhf=96\gev$,
the efficiency for selecting four photons is significantly enhanced, about 40\% for BP3 and about 27\% for BP4.
The mass of $\hf$ is heavy enough to produce a pair of hard photons.
If we include the events with $n_\gm \geq 3$,
the selection efficiency for $\mhf=96\gev$ is more than 70\% for BP3 and 60\% for BP4.
Then even BP3 with the signal rate of $\mco(0.1)\fb$
can yield a few hundred events with the expected luminosity of $3\ab^{-1}$.
Considering almost background-free environment of $4\gm+VV'$,
we expect that the HL-LHC can probe the fermiophobic type-I.

  \section{Conclusions}
 \label{sec:conclusions}
 
The null results in the BSM searches at the LHC can coexist with an intermediate-mass new particle 
if the new particle is fermiophobic.
Type-I in the 2HDM with the condition of $\al=\pi/2$ provides a fermiophobic Higgs boson,
a lighter \textit{CP}-even Higgs boson.
The heavier \textit{CP}-even Higgs boson describes the observed SM-like Higgs boson at a mass of $125\gev$.
In this letter, we have studied the validity of the fermiophobic type-I
not only at the electroweak scale but also at higher energy scale. 

For the comprehensive study,
we have first obtained the still-valid parameter space satisfying the theoretical requirements 
(bounded-from-below scalar potential, unitarity, perturbativity, vacuum stability),
flavor-changing neutral currents in $B$ physics, the cutoff scale above $1\tev$, Higgs precision data, 
and direct collider search bounds from the LEP, Tevatron, and LHC.
For every viable parameter point, the cutoff scale $\lmc$ has been calculated
through the RGE analysis.
The most important result is that the fermiophobic type-I
can maintain the theoretical validity all the way up to the Planck scale when the mass
of the fermiophobic Higgs boson is larger than about $62.5\gev$.
Even the mild condition of $\lmc>1\tev$ severely restricts the parameter space,
which predicts definite signal rates of the BSM Higgs bosons in the multi-photon states at the LHC.
Based on the studies of the decay rates and the production cross sections of the BSM Higgs bosons,
we have suggested the $4\gm+VV'$ processes as the golden discovery channels for the model
since they enjoy almost background-free environment and substantial selection efficiencies.

\acknowledgments
The work of JK, SL, and JS is supported by 
the National Research Foundation of Korea, Grant No.~NRF-2022R1A2C1007583. 
The work of P.S. was supported by the appointment to the JRG Program at the APCTP through the Science and Technology Promotion Fund and Lottery Fund of the Korean Government. This was also supported by the Korean Local Governments - Gyeongsangbuk-do Province and Pohang City.

\bibliographystyle{JHEP}
\bibliography{./fphobic-4aVV}

\end{document}